# Global versus Local Weak-Indication Self-Timed Function Blocks – A Comparative Analysis

P. BALASUBRAMANIAN*, N.E. MASTORAKIS[§][¶]
* School of Computer Engineering
Nanyang Technological University
50 Nanyang Avenue
SINGAPORE 639798
Email: balasubramanian@ntu.edu.sg
[§] Department of Computer Science
Military Institutes of University Education
Hellenic Naval Academy
Piraeus 18539, GREECE
Email: mastor@hna.gr
[¶] Department of Industrial Engineering
Technical University of Sofia
Sofia 1000, Boulevard Kliment Ohridski 8
BULGARIA
Email: mastor@tu-sofia.bg

*Abstract:* - This paper analyzes the merits and demerits of global weak-indication self-timed function blocks versus local weak-indication self-timed function blocks, implemented using a delay-insensitive data code and adhering to 4-phase return-to-zero handshaking. A self-timed ripple carry adder is considered as an example function block for the analysis. The analysis shows that while global weak-indication could help in optimizing the power, latency and area parameters, local weak-indication facilitates the optimum performance in terms of realizing the data-dependent cycle time that is characteristic of a weak-indication self-timed design.

## 1 Introduction

The International Technology Roadmap for Semiconductors (ITRS) [1] has identified design for reliability and resilience as one of the futuristic grand challenges for design technology in its 2011 edition. The percentage of design reuse in system-on-chip designs which was estimated to be 46% in the 2009 ITRS edition is expected to become 96% by the year 2024. Further, the proportion of design blocks reuse with respect to glue logic is expected to reach 60% in the year 2024. Moreover, parameter uncertainty as a percentage effect on sign-off delay is expected to increase from a current level of 18% to 32% by 2024. In this backdrop, self-timed design, which is a robust flavor of asynchronous design, is acclaimed to be a strong contender or an inevitable counterpart for digital circuit/system designs in the nanoelectronics regime. This is because self-timed designs are inherently modular (i.e. permits design reuse) [2], are self-checking [3], exhibit superior EMI compatibility [4], are noise-tolerant [5], have the ability to cope with parametric uncertainty, supply voltage, threshold voltage, and temperature variations [6] [7], consume power only when and where active [8], and are able to guarantee greater security and robustness against hostile attacks compared to synchronous designs in the case of sensitive industrial applications [9] [10]. Taking cognizance of these facts, the ITRS design report has projected a growing requirement for asynchronous signaling in the nanoelectronics era and also emphasizes the continuous development of asynchronous circuit and system design tools over the foreseeable future.

In this paper, we focus on analyzing the merits and demerits of global weak-indication self-timed function blocks versus local weak-indication self-timed function blocks by building upon a prior work [11], which reported that global weak-indication is preferable compared to local weak-indication for realizing a cascade of function blocks from power, latency and area perspectives. Whilst confirming that the observations reported in the previous work [11] are correct, we additionally show that local weak-indication is actually preferable from the point of view of cycle time than global weak-indication. In other words, we clarify that global weak-indication could reduce the throughput aspect of a weak-indication self-timed design, and with respect to realizing the true timing advantage inherent in a weak-indication self-timed design, the local weak-indication design method is indeed preferable.

In the remainder of this paper, Section 2 presents the fundamental concepts of self-timed design with some illustrations. Section 3 briefly discusses the local and global weak-indication self-timed system architectures by considering a 32-bit ripple carry adder (RCA) as an example function block. This is followed by the simulation results obtained for two local and global weak-indication self-timed 32-bit RCAs. Subsequently, the theoretical evaluation of cycle times for the two RCAs corresponding to local and global weak-indication is presented. Finally, the conclusions are derived in Section 4.

## 2 Preliminaries and Background

A self-timed (i.e. asynchronous) function block is the combinational logic equivalent of a synchronous digital system [12] [13]. Self-timed function blocks represent a robust flavor of asynchronous function blocks and are constructed using delay-insensitive data codes which adhere to 4-phase return-to-zero (RTZ) handshaking.

The dual-rail code is the simplest member of the generic family of delay-insensitive data codes [14], based on which a data wire X is represented using the dual wires X1 and X0, as shown in Fig. 1. X = 1 is represented by X1 = 1 and X0 = 0, and X = 0 is represented by X1 = 0 and X0 = 1. These two conditions represent 'valid data', and the condition of both X1 and X0 assuming 0 is referred to as the 'spacer'. The 4-phase RTZ handshaking procedure requires that the application of inputs from the external environment follows the defined sequence: valid data-spacer-valid data-spacer, and so forth.

The representative block diagram of a typical self-timed system stage is shown in Fig. 1 that is accompanied by the sender-receiver analogy. The valid data and spacer processing operations of a self-timed system stage have been explained in [11] [12] [13], and the reader is referred to the same. In Fig. 1, the junction dots shown enclosed within the pink ovals in dotted lines represents isochronic fork junctions. Isochronicity constitutes the weakest compromise to delay-insensitivity [15], and an isochronic fork junction implies that all the nets forking out from the same junction experience similar signal transitions occurring simultaneously.

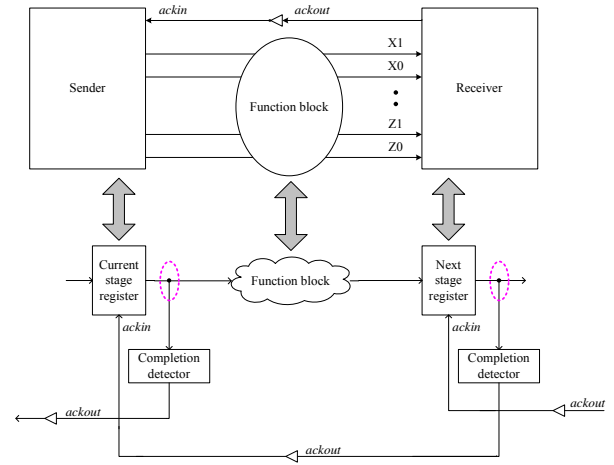

Fig. 1 Block diagram of a self-timed system stage

Referring to Fig. 1, the 4-phase RTZ protocol is explained as follows. The dual-rail data bus that feeds the current stage register (i.e. sender) is initially in the spacer state, and the common acknowledge input (*ackin*) for the current stage register is binary 1, since the common acknowledge output (*ackout*) provided by the next stage register (i.e. receiver) is binary 0. The current stage register now transmits a codeword (i.e. valid data). This results in low to high transitions on anyone of the corresponding rails of all the dual-rail bus wires which feed the function block. After the next stage register receives a codeword subsequent to completion of data processing in the function block, it drives the *ackout* wire to 1, and the *ackin* wire assumes 0. The current stage register waits for the *ackin* signal to become 0 and then resets the data bus, i.e. the data bus feeding the function block is driven to the spacer state. After an unbounded but finite and positive amount of time taken for the resetting of the function block and the passage of the spacer to the following register stage, the next stage register drives the *ackout* (*ackin*) to 0 (1). With this, a single data transaction is said to be complete, and the system is ready to commence the next transaction.

Self-timed function blocks are primarily classified as strongly indicating [16] [17], weakly indicating [16] [18], and early output type [19] [20]. Indication means acknowledging the receipt of primary inputs through the intermediate/primary outputs. The input-output timing relationship of strong-indication, weak-indication, and early output type self-timed function blocks is illustrated by a representative timing diagram shown as Fig. 2.

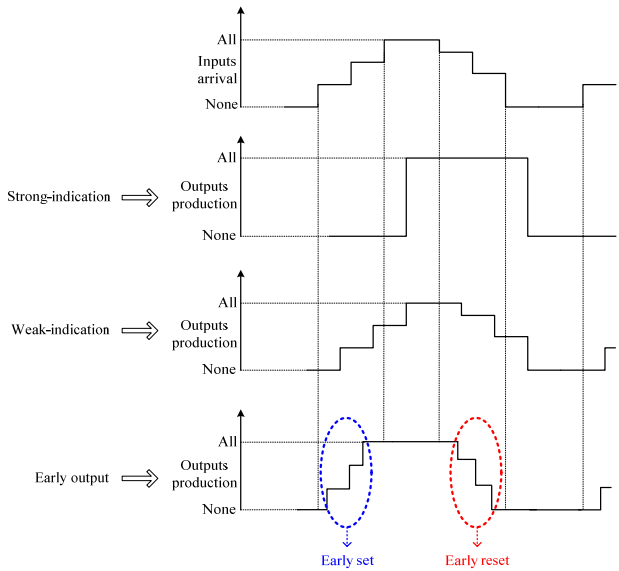

Fig. 2 Input-output timing correlation of strong-indication, weak-indication, and early output type function blocks

A strong-indication function block starts to produce the primary outputs only after receiving all the primary inputs whether they are valid data or spacers. A weak-indication function block is able to produce the primary outputs after receiving a subset of the primary inputs. However, the production of at least one primary output is withheld till all the primary inputs are received. The early output type function block is more relaxed compared to strong and weak-indication function blocks in that it could produce all the primary outputs after receiving just a subset of the primary inputs. The early output function block is further subdivided into 2 types as early set and early reset. The early set type function block is capable of producing all the valid primary outputs after receiving a subset of the valid primary inputs. On the other hand, the early reset type function block is capable of producing all the spacer primary outputs after receiving a subset of the spacer primary inputs. The early output function block would be of either early set or early reset type, and their respective behaviors are portrayed by the blue and red ovals shown in dotted lines in Fig. 2.

## 2.1 Strong-Indication Full Adder Block

The strong-indication full adder derived on the basis of the delay-insensitive minterm synthesis (DIMS) method [21] is portrayed by Fig. 3. Here, (A1, A0), (B1, B0) and (CIN1, CIN0) represent the dual-rail primary inputs viz. augend, addend and carry inputs of the full adder respectively, and (SUM1, SUM0), and (COUT1, COUT0) represent the corresponding dual-rail sum and carry outputs of the full adder. The primary input and output labels shall be uniformly maintained for all the full adders which will be discussed in this work. The logic equations governing the DIMS strong-indication full adder are given below. It can be noticed in (1) to (4) that the product terms comprising the dual-rail sum and carry output expressions are mutually disjoint [22] [23] [31], i.e. the logical conjunction of any pair of product terms would result in null.

$$SUM1 = A0B0CIN1 + A0B1CIN0 + A1B0CIN0 + A1B1CIN1 \qquad (1)$$

$$SUM0 = A0B0CIN0 + A0B1CIN1 + A1B0CIN1 + A1B1CIN0 \qquad (2)$$

$$COUT1 = A0B1CIN1 + A1B0CIN1 + A1B1CIN0 + A1B1CIN1 \qquad (3)$$

$$COUT0 = A0B0CIN0 + A0B0CIN1 + A0B1CIN0 + A1B0CIN0 \qquad (4)$$

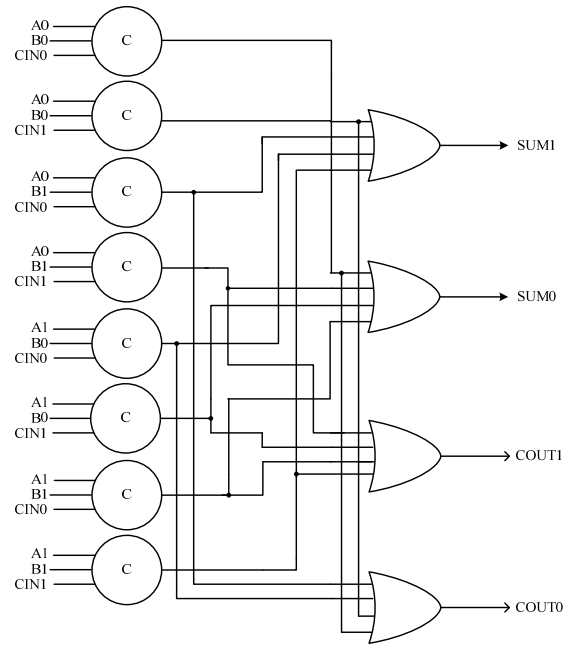

Fig. 3 DIMS strong-indication full adder

It can be seen from (1) to (4) that the dual-rail sum and carry outputs are dependent upon all the primary inputs, thus satisfying the strong-indication timing constraints. The products are implemented using C-elements, which are represented by the circles with the marking 'C' on their periphery in the figures. An *n*-input C-element will output 1 only when all its *n*-inputs are 1. Similarly, an *n*-input C-element will output 0 only when all its *n*-inputs are 0. Even if a single input of an *n*-input C-element assumes a different binary value, then the C-element will just hold on to its existing steady-state, i.e. there will not be any change in the C-element's output.

## 2.2 Weak-Indication Full Adder Blocks
Reference [24] discusses three kinds of weak-indication full adder realizations – basic, distributive, and biased.

### 2.2.1 Basic Implementation
A basic weak-indication full adder realization corresponding to the DIMS approach [21] is shown in Fig. 4. The dual-rail sum output expressions of this full adder are identical to (1) and (2). However, the dual-carry output expressions are different from (3) and (4) and are given by (5) and (6) below. Nevertheless, the product terms comprising the dual-rail carry output are mutually disjoint.

$$COUT1 = A0B1CIN1 + A1B0CIN1 + A1B1 \quad (5)$$

$$COUT0 = A0B1CIN0 + A1B0CIN0 + A0B0 \quad (6)$$

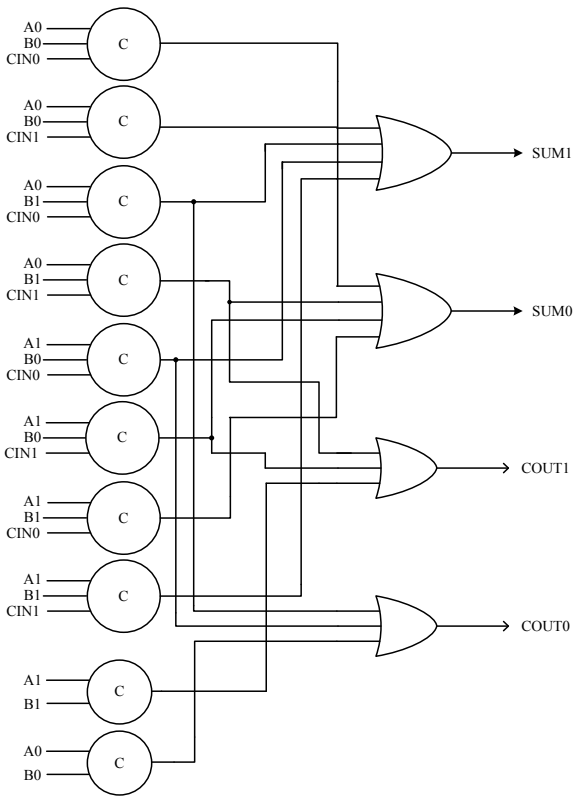

Fig. 4 DIMS weak-indication full adder

From (5) and (6), it can be noted that while the dual-rail sum output depends upon all the primary inputs, the dual-rail carry output need not. Thus the indication of all the primary inputs is provided by the sum output. The dual-rail carry output may take advantage of the carry-generate (i.e. A1=B1=1) or the carry-kill condition (i.e. A0=B0=1), rather than the slow carry-propagate condition which is dictated by A0=B1=1 or A1=B0=1. Since the carry output can be generated or killed by a full adder regardless of the incoming carry (CIN1/CIN0) when suitable augend and addend inputs are supplied, and because the sum output would take care of acknowledging the receipt of all the primary inputs, the full adder shown in Fig. 4 is said to be weakly indicating. Notice that when the carry-propagate condition becomes active, dual acknowledgments of the primary inputs would result since the dual-rail sum and carry outputs would indicate the receipt of all the dual-rail primary inputs.

### 2.2.2 Distributive Implementation
The weak-indication full adder [25] corresponding to the distributive implementation style [26] is depicted by Fig. 5. This full adder implements (7) to (10), which are obtained through logic factorization [27] of (1), (2), (5) and (6) respectively.

$$SUM1 = (A0B0 + A1B1)\, CIN1 \\ + (A0B1 + A1B0)\, CIN0 \quad (7)$$

$$SUM0 = (A0B0 + A1B1)\, CIN0 \\ + (A0B1 + A1B0)\, CIN1 \quad (8)$$

$$COUT1 = (A0B1 + A1B0)\, CIN1 + A1B1 \quad (9)$$

$$COUT0 = (A0B1 + A1B0)\, CIN0 + A0B0 \quad (10)$$

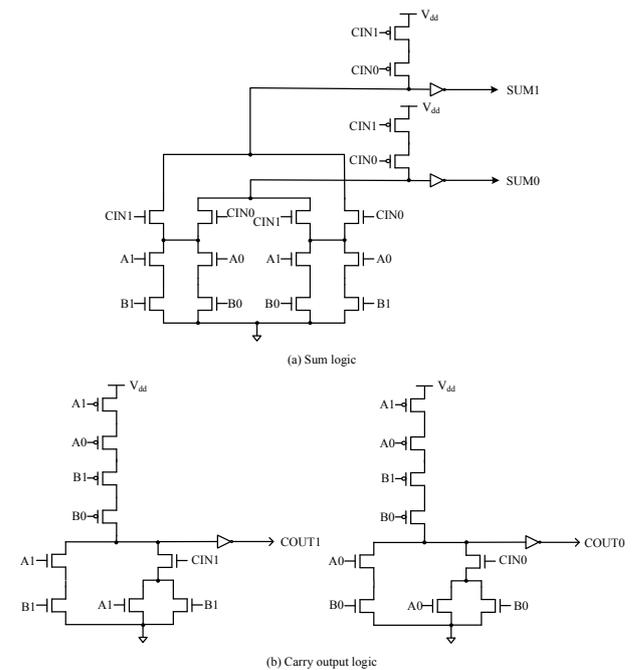

Fig. 5 Martin's full adder

Martin's full adder [25], shown in Fig. 5, is an optimized full-custom transistor level implementation of (7) to (10). Equations (7) and (8) show that the dual-rail sum output depends upon all the dual-rail primary inputs, while (9) and (10) show that the dual-rail carry output may not have to depend on the dual-rail carry input always, which is characteristic of the weak-indication timing model.

For application of valid data, the dual-rail sum output is produced based on all the dual-rail primary inputs, and the dual-rail carry output may be produced based on only the augend and addend primary inputs if carry-generate or carry-kill condition occurs, and inclusive of the carry input if and only if carry propagation occurs. During RTZ, however, the dual-rail carry output would be reset after the reset of the augend and addend primary inputs in the pMOS network, and the RTZ of the incoming carry is indicated only by the dual-rail sum output through its pMOS network. Thus the responsibility of indicating the RTZ of dual-rail augend, addend, and carry inputs is distributed between the dual-rail sum and carry outputs during the RTZ phase. Hence Martin's full adder is said to correspond to the distributive weak-indication implementation style.

### 2.2.3 Biased Implementation

The self-timed full adder proposed in [28] that corresponds to the biased weak-indication implementation style is portrayed by Fig. 6. It implements the dual-rail sum output given by (7) and (8), while the dual-rail carry output equations which correspond to majority logic are given below.

$$COUT1 = A1B1 + B1CIN1 + A1CIN1 \qquad (11)$$

$$COUT0 = A0B0 + B0CIN0 + A0CIN0 \qquad (12)$$

Both (11) and (12) are implemented using a single complex gate viz. the AO222 cell as shown in Fig. 6. Unlike the C-element, the AO222 cell is an input-incomplete gate. The C-element waits for all the inputs to assume the same binary state to reflect it on the output – hence it is strongly indicating and input-complete. On the contrary, the AO222 cell is able to output 1 even when one of its constituent product terms formed using just 2 out of 3 literals is activated, i.e. for application of valid data, the activation of any product term(s) would result in the production of 1 through the AO222 cell. For application of spacer in the subsequent RTZ phase, even if anyone literal of the product term(s) which was previously activated now becomes 0, the AO222 would output 0. These two conditions imply that the AO222 cell is non-indicating unlike the Muller C-element. In Fig. 6, the dual-rail sum output is dependent upon all the dual-rail primary inputs, while the dual-rail carry output is completely freed from indication. Thus there is a bias towards the carry output in contrast with the sum output, and hence the full adder in Fig. 6 is said to correspond to the biased weak-indication implementation style.

In the case of the basic weak-indication full adder shown in Fig. 4, when carry propagation occurs, the receipt of the dual-rail primary inputs is doubly acknowledged by the dual-rail sum and carry outputs. When carry-generate or carry-kill occurs, the receipt of the dual-rail augend and addend inputs are doubly acknowledged by the dual-rail sum and carry outputs, but the receipt of the carry input is only acknowledged by the sum output. On the contrary, when carry propagation occurs in Fig. 6, the receipt of the dual-rail primary inputs is duly acknowledged by the dual-rail sum output alone. The dual-rail carry output would not be able to provide an acknowledgment since the carry output logic is synthesized using input-incomplete gates. This is the same scenario for carry generation and carry kill as well. Also, unlike Martin's full adder, the responsibility of indicating the primary inputs is not distributed between the dual-rail sum and carry outputs of the full adder shown in Fig. 6.

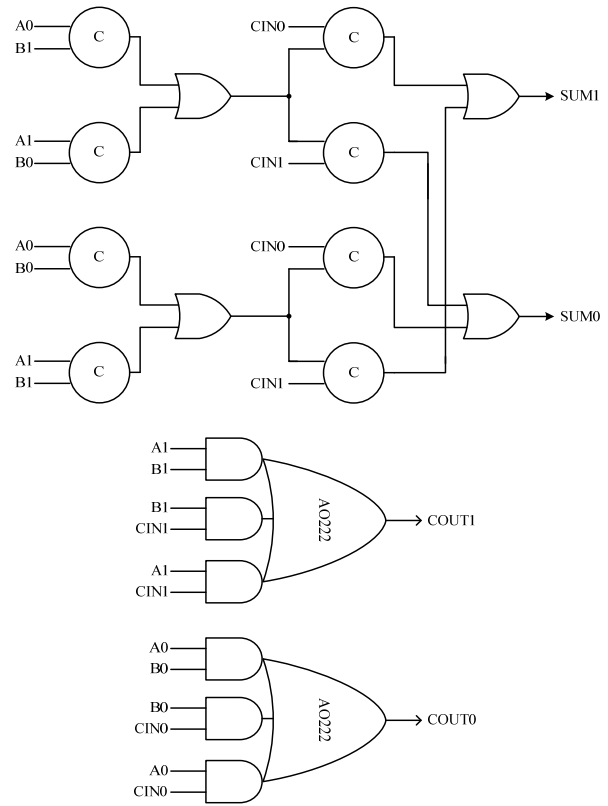

Fig. 6 Biased weak-indication full adder

## 2.3 Early Output Full Adder Block

A dual-rail full adder [20] that corresponds to the early output (here, early reset) logic [19] is shown in Fig. 7, which synthesizes (7) to (10).

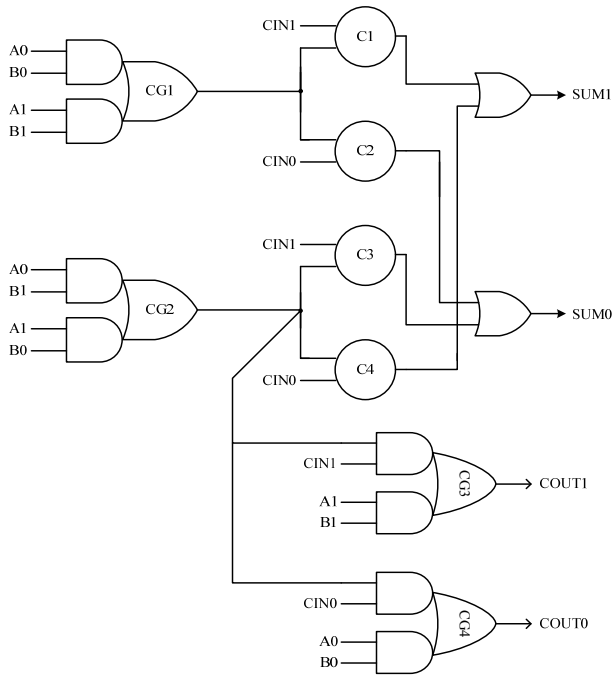

Fig. 7 Early output full adder

The early output full adder shown above contains eight complex gates and two simple gates. The complex gates are marked as CG1 to CG4 and C1 to C4 in Fig. 7. C1 to C4 are 2-input C-elements. The 2-input C-element is realized using an AO222 gate with feedback, and implements the logic function Z = XY+(X+Y) Z, where X and Y represent the inputs and Z represents the output. Gates CG1 to CG4 are AO22 cells. An AO22 complex gate with inputs A, B, C and D and output Y implements the Boolean function Y = AB + CD. Note that excepting the C-gate and the inverter, all other logic gates are input-incomplete gates. It may be noted that logic redundancy is implicit [29] in the full adder shown in Fig. 7.

Let us consider three example input scenarios viz. carry propagation, carry generation, and carry-kill to succinctly describe the early output viz. early reset operation of the full adder shown in Fig. 7.

- *Carry propagation*: A0=B1=1 or A1=B0=1

For application of this valid data, the complex gate CG2 will be activated and its output will become 1. Assuming CIN1 to be 1, the C-element C3 will be activated and SUM0 will attain 1. Since CIN1 is presumed to be 1, the complex gate CG3 will be activated and hence COUT1 will attain 1. In a subsequent RTZ phase, even with A0 or B1 (or A1 or B0) becoming spacer, COUT1 could assume the spacer state irrespective of CIN1 attaining the spacer. Also, with A0 or B1 (or A1 or B0) assuming the spacer, and with CIN1 assuming the spacer, SUM0 could also assume the spacer state. Thus even with any two out of three primary inputs attaining the spacer, i.e. A0 or B1 (or A1 or B0) and CIN1 assuming the spacer, the entire full adder could be reset, which is reflective of its early reset nature.

- *Carry generation*: A1=B1=1

In this case, for application of valid data, the complex gate CG1 will be activated and its output will become 1. Under this condition, if CIN1 is 1, the C-element C1 will be activated and its output will become 1. Eventually, SUM1 will attain 1. Since A1 and B1 are 1, the complex gate CG3 will be activated and hence COUT1 will also attain 1. In a subsequent RTZ phase, even with A1 or B1 becoming spacer, COUT1 will become 0. Likewise, even with A1 or B1 becoming spacer, the output of gate CG1 will assume 0, and with CIN1 also becoming 0, the output of C1 will become 0 and SUM1 will attain 0. Thus with either A1 or B1 attaining 0 in the RTZ phase and accompanied by the RTZ of CIN1, COUT1 and SUM1 will become 0. This again demonstrates the early reset nature of the full adder.

- *Carry-kill*: A0=B0=1

In this case, for application of valid data, the complex gate CG1 will be activated and its output will become 1. Under this condition, if CIN0 is 1, the C-element C2 will be activated and its output will become 1. Eventually, SUM0 will attain 1. Since A0 and B0 are 1, the complex gate CG4 will be activated and hence COUT0 will attain 1. In a subsequent RTZ phase, even with A0 or B0 becoming spacer, COUT0 will become 0. Likewise, even with A0 or B0 becoming spacer, the output of gate CG1 will assume 0, and with CIN0 also becoming 0, the output of C2 will become 0 and SUM0 will attain 0. Thus with either A0 or B0 attaining 0 in the RTZ phase and accompanied by the RTZ of CIN0, COUT0 and SUM0 will become 0, which once again demonstrates the early reset property of the full adder.

## 2.4 Self-Timed RCAs – Timing Attributes

It is important to note that when small function blocks are cascaded to form a large function block, the respective timing property is preserved [16]. For example, cascading of strong-indication full adders results in a strong-indication RCA. Strong-indication, weak-indication, and early output $n$-bit self-timed RCAs would be characterized by the generalized timing attributes given in Table 1 [24] [30].

Table 1: Latencies and cycle time magnitudes of strong-indication, weak-indication, and early output $n$-bit RCAs; $m$ denotes carry propagation length

| RCA realization style | Forward latency | Reverse latency | Cycle time |
|---|---|---|---|
| Strong-indication | $O(n)$ | $O(n)$ | $O(2n)$ |
| Weak-indication (Basic) | $O(m)$ | $O(m)$ | $O(2m)$ |
| Weak-indication (Distributive/ Biased) | $O(m)$ | $O(2)$ | $O(m+2)$ |
| Early output | $O(m)$ | $O(2)$ | $O(m+2)$ |

Forward latency is the time taken for processing of valid data by a self-timed system stage, and reverse latency is the time encountered in resetting a self-timed system stage, i.e. the time taken for spacer processing by a self-timed system stage. The metric 'cycle time' specifies the sum of forward and reverse latencies. Cycle time is the metric that quantifies the timing performance viz. throughput of a self-timed design since the cycle time denotes the time required to complete a single data transaction, i.e. processing of valid data followed by RTZ. From Table 1, it can be seen that the distributive/biased weak-indication and the early output $n$-bit RCAs feature similar optimized cycle time compared to the other self-timed RCAs.

## 3 Local and Global Weak-Indication Self-Timed System Architectures

The architectures of local and a global weak-indication self-timed system stage are depicted by Fig. 8a and 8b respectively. The isochronic regions are highlighted by the pink ovals in dotted lines.

Notice that Fig. 8a is identical to Fig. 1 described earlier, while Fig. 8b is very similar to Fig. 8a with the inclusion of the synchronizer. The synchronizer basically consists of synchronization elements viz. 2-input C-elements. At least, one dual-rail primary output of the early output function block in Fig. 8b is synchronized with the completion detector output (i.e. *ackout*), and the outputs of the synchronizer are supplied to the next stage besides the other primary outputs of the early output function block which are directly supplied to the next stage. In contrast, all the primary outputs of the weak-indication function block in Fig. 8a are directly supplied to the next stage.

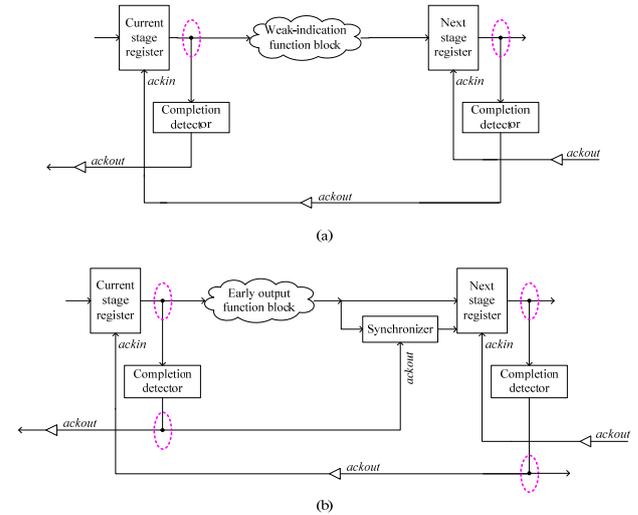

fig. 8 (a) Local weak-indication system stage and (b) Global weak-indication system stage

In the case of Fig. 8a, after receiving even a subset of the primary inputs, the weak-indication function block could process them to produce the primary outputs. But only after all the primary inputs are received by the weak-indication function block, the production of all the primary outputs becomes complete which are directly delivered to the next stage. However, in the case of Fig. 8b, after receiving just a subset of the primary inputs, the early output function block may process them and in turn produce all the primary outputs. But since one dual-rail primary output is given to the synchronizer, the synchronizer waits to perform synchronization of the dual-rail primary output with the completion detector output (*ackout*). Receipt of *ackout* by the synchronizer confirms that all the primary inputs have been received by the early output function block due to isochronic assumptions imposed on all the primary input forks. Hence, after ensuring synchronization, the synchronizer delivers the withheld dual-rail primary output to the next stage. Therefore the synchronizer governs the responsibility of satisfying the weak-indication criteria globally with respect to the self-timed system stage shown in Fig. 8b contrary to locally satisfying the weak-indication criteria in Fig. 8a.

For simulation purpose, it is presumed that the function blocks in Fig. 8a and 8b are represented by 32-bit RCAs, formed by the cascade of 32 numbers of a full adder block. Cascading of weak-indication full adders, such as those illustrated in Section 2.2, gives rise to a weak-indication RCA, and cascading of early output full adder blocks, such as the one illustrated in Section 2.3, results in an early output RCA. With reference to Fig. 8b, the 32 dual-rail sum outputs of the early output RCA are directly supplied to the next stage, while the dual-rail carry overflow output of the RCA alone is supplied to the next stage through the synchronizer after a rendezvous with the completion detector output.

Sample simulation results corresponding to local and global weak-indication self-timed function blocks (here, 32-bit RCAs) are given in Table 2. For realizing the local weak-indication 32-bit RCA, the weakly indicating full adder proposed in [24] was used, and for realizing the global weak-indication 32-bit RCA, the early output full adder discussed in Section 2.3 was used. The simulation results correspond to a 32/28nm CMOS process [32].

Table 2: Simulation results obtained for local and global weak-indication self-timed 32-bit RCAs based on a 32/28nm CMOS process

| Function block Type | Power ($\mu$W) | Latency (ns) | Area ($\mu m^2$) |
|---|---|---|---|
| Local weak-indication | 2171 | 3.32 | 2049.16 |
| Global weak-indication | 2161 | 3.10 | 1665.41 |

The local and global weak-indication self-timed 32-bit RCAs were constructed in a semi-custom ASIC-based design style using standard cells. The structural integrity of the full adders and the RCAs were preserved during physical realization viz. technology mapping which paves the way for a straightforward comparison of the design metrics post-physical synthesis. The 2-input C-element was alone designed manually using the AO222 cell functionality by incorporating feedback and was made available to realize the various self-timed designs. The 3-input C-elements were decomposed into 2-input C-elements wherever necessitated using the safe QDI logic decomposition method of [33]. Unsafe logic decomposition could lead to gate or wire orphans [34] [35], where an orphan refers to an unacknowledged signal transition.

A self-timed RCA stage comprises the function block, the input registers, and the completion detector. The input registers and the completion detector part of the local and global weak-indication self-timed RCAs are identical, and only their function blocks and the synchronizer logic differ. Hence the differences between the simulations results of the two RCAs can be attributed primarily to the differences between their constituent function blocks and secondarily to the synchronizer present only in case of the global weak-indication self-timed RCA. More than 1000 random input vectors were supplied to the two RCAs at time intervals of 20ns through identical test benches in order to capture the switching activities. The value change dump files generated during the simulations were subsequently used for power estimation using Synopsys tool. Since the EDA tool mainly estimates critical path timing, the worst-case forward latency was alone estimated for a typical case PVT specification. Appropriate wire loads (i.e. parasitic) were included whilst performing the simulations. As part of advanced timing analysis, a virtual clock was used to constrain the input and output ports of the RCAs, and it did not contribute to any power dissipation.

From the simulation results given in Table 2, it is clear that global weak-indication enables reduction in latency by 6.6% and reduction in area by 18.7% compared to local weak-indication without any power penalty. The areas of the weak-indication full adder and the early output full adder used in local and global weak-indication self-timed RCAs are 39.65$\mu m^2$ and 27.45$\mu m^2$ respectively – the lesser area of the latter is attributable to the less number of logic gates and the less number of complex logic gates present in it compared to the former. The synchronizer area corresponding to global weak-indication is 6.61$\mu m^2$.

Although global weak-indication facilitates reductions in latency and area without increasing the power dissipation, local weak-indication benefits in terms of cycle time, which governs the throughput of a self-timed design. To elaborate on this, a theoretical evaluation of the cycle times pertaining to local and global weak-indication self-timed systems are discussed by assuming the function block to be a 32-bit self-timed RCA. To aid with this, the details of a generic self-timed system stage is given in Fig. 9 where the circuit portions enclosed within the dotted rectangles represent the internal details. The data path logic and the synchronizing logic are highlighted in blue and green respectively. The portion portrayed in red corresponds to global weak-indication. Exclusion of the portion portrayed in red, accompanied by a direct forwarding of all the dual-rail sum and carry outputs of the RCA to the next stage is characteristic of local weak-indication. The completion detector circuit detail is also given along with a depiction of the number of logic levels.

### 3.1 Cycle Time Evaluation for Local Weak-Indication Self-Timed System

The data path logic, highlighted in blue in Fig. 9, comprises the current stage register and the self-timed RCA. The input data is supplied through the current stage register to the self-timed RCA for data processing. The current and next stage registers are made up of a bank of 2-input C-elements, with a 2-input C-element dedicated for each dual-rail input that is coupled with the *ackin* signal.

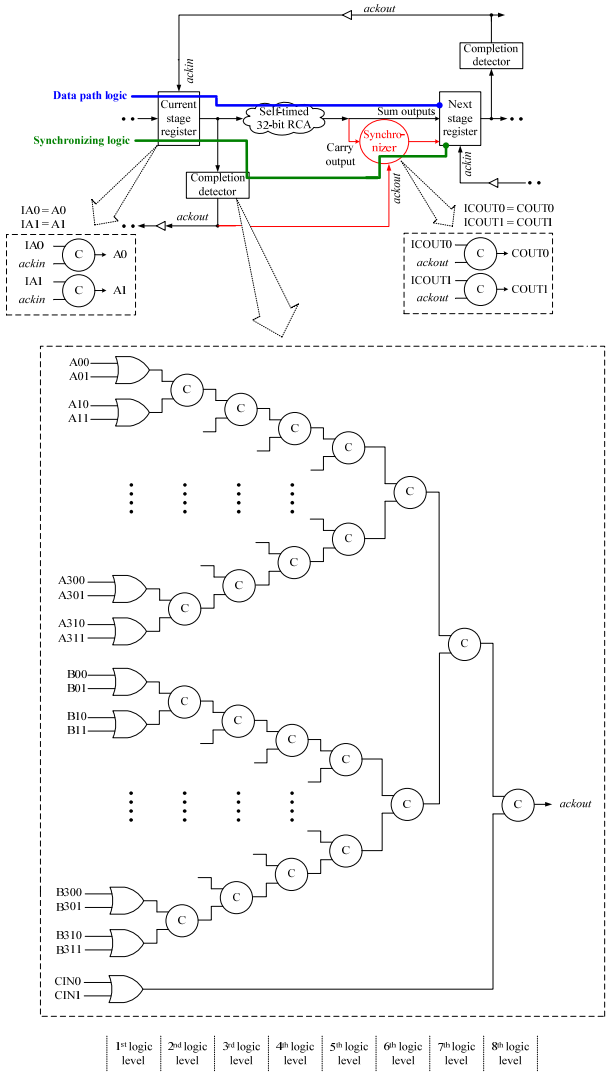

Fig. 9 Details of self-timed system architecture stage
Note: Exclusion of the portion highlighted in red and direct forwarding of the sum and carry outputs of the self-timed RCA to the next stage signifies local weak-indication, while inclusion of the portion highlighted in red signifies global weak-indication

The worst-case data path logic delay (i.e. forward latency) encountered for processing of valid data is given by (13) and the worst-case data path logic delay (reverse latency) encountered for processing of spacer is given by (14). Note that (14) holds well only when a distributive or biased weak-indication full adder is duplicated to form a self-timed RCA [24]. On the other hand, if a basic weak-indication full adder is duplicated to form an RCA, its reverse latency would be specified by (13) instead of (14).

The term T is used to denote propagation delay in the equations, with 'ns' being the unit. $T^{VD}$ represents the propagation delay encountered for processing valid data, and $T^S$ represents the propagation delay encountered for spacer processing. $T_{gate}$ refers to the propagation delay of a specific gate. In (13), on the right hand side, the first term given within brackets signifies the current stage register delay; the second term given within brackets specifies the delay associated with carry output production in the least significant full adder stage; the third term given within brackets signifies the delay associated with carry propagation through $m$ arbitrary full adder stages within the RCA where $m < n$, and $n$ represents the RCA size (here, $n$ = 32). Lastly, the fourth term given within brackets on the right hand side denotes the delay associated with the most significant sum output production.

$$T^{VD}_{\text{data path logic}} = T_{CE2} + (T_{CE2} + T_{OR2} + T_{AO21}) \\ + (T_{AO21} \times m) + (T_{CE2} + T_{OR2}) \quad (13)$$

$$T^{S}_{\text{data path logic}} = T_{CE2} + (T_{CE2} + T_{OR2} + T_{AO21}) \\ + (T_{CE2} + T_{OR2}) \quad (14)$$

It may be deduced from (13) and (14) that $T^S$ is constant and $T^{VD} \geq T^S$. The cycle time is given by,

$$T^{local}_{cycle} = 6T_{CE2} + 4T_{OR2} + (m + 2) \times T_{AO21} \quad (15)$$

### 3.2 Cycle Time Evaluation for Global Weak-Indication Self-Timed System

The synchronizing logic, highlighted in green in Fig. 9, consists of the current stage register, the completion detector, and the synchronizer. The outputs of data path logic and synchronizing logic terminate at the next stage register, serving as its inputs. In the case of global weak-indication, the forward (reverse) latency would be dictated by the maximum propagation delay associated with the data path or synchronizing logic, whichever is higher, for application of valid data (spacer). Two possible scenarios may arise: (i) only carry-generate or carry-kill occurs and dominates carry propagation within the RCA in terms of the number of stages – in either of these cases, the synchronizing logic delay may dominate the data path logic delay, and (ii) carry-propagate occurs and dominates carry generation or carry kill occurring within the RCA in terms of the number of stages – in this case, the data path logic delay may be equal to or may exceed the synchronizing logic delay.

The data path logic delay involved in processing valid data is given by (16) and that for spacer processing is given by (17).

$$T^{VD}_{data\ path\ logic} = T_{CE2} + 2T_{AO22} + (T_{AO22} \times m) + (T_{CE2} + T_{OR2}) \quad (16)$$

$$T^{S}_{data\ path\ logic} = T_{CE2} + 2T_{AO22} + (T_{CE2} + T_{OR2}) \quad (17)$$

The synchronizing logic delay is given by,

$$T_{synchronizing\ logic} = T_{CE2} + T_{OR2} + 7T_{CE2} + T_{CE2} \quad (18)$$

In (18), on the right hand side, the first term signifies the current stage register delay, the second term represents the delay associated with the 2-input OR gate present in the first logic level of the completion detector shown in Fig. 9, the third term denotes the combined delay of seven 2-input C-elements present in the second to eighth logic levels of the completion detector shown in Fig. 9, and the final term in (18) signifies the synchronizer delay.

If significant carry propagation occurs within the RCA when valid data is applied, then $T^{VD}_{data\ path\ logic}$ may be greater than or equal to $T_{synchronizing\ logic}$. However, when spacer data is applied in the subsequent RTZ phase, $T_{synchronizing\ logic}$ may well dominate $T^{S}_{data\ path\ logic}$. In other words, a fast RTZ that is bound by a constant latency as specified by (14) which is realizable in the local weak-indication RCA may not be realized in the global weak-indication RCA since $T_{synchronizing\ logic}$ of (18) tends to dominate $T^{S}_{data\ path\ logic}$ given by (17). Hence the early reset nature of the full adders constituting the global weak-indication RCA does not benefit in terms of cycle time reduction. The cycle time of the global weak-indication RCA is therefore given by,

$$T^{global}_{cycle} = 11T_{CE2} + 2T_{OR2} + (m + 2) \times T_{AO22} \quad (19)$$

Equating (16) and (18) and substituting the typical propagation delay values of minimum sized gates present in the library [32], and solving for $m$ gives,

$$m \leq 8 \quad (20)$$

The inequality mentioned in (20) implies that $T_{synchronizing\ logic}$ will dominate $T^{VD}_{data\ path\ logic}$ while processing valid data if the worst-case carry propagation within the global weak-indication RCA is limited to a maximum of 8 full adder stages, exceeding which $T^{VD}_{data\ path\ logic}$ will dominate $T_{synchronizing\ logic}$. Provided (20) holds well, the cycle time of the global weak-indication RCA may be expressed as,

$$T^{global}_{cycle} = 18T_{CE2} + 2T_{OR2} \quad (21)$$

Substituting the typical propagation delay values of minimum sized gates present in the library [32], in (15) and (19), we get,

$$T^{local}_{cycle} = 63m + 1002 \quad (22)$$

$$T^{global}_{cycle} = 72m + 1430 \quad (23)$$

A plot of equations (22) and (23) for different carry propagation lengths (*m*) varying from 4-bits to 28-bits is portrayed through Fig. 10. It is evident from Fig. 10 that the local weak-indication self-timed system consistently outperforms the global weak-indication self-timed system in terms of throughput viz. cycle time, although the latter was reported to be better than the former with respect to optimization of design parameters such as power, latency and area (refer to Table 2). A similar inference is likely to be made when considering local and global weak-indication self-timed RCAs constructed using a cascade of hybrid input encoded full adders [36] or homogeneous or heterogeneously encoded dual-bit full adder blocks [37] [38].

Based on equations (22) and (23) and the cell library information given in [32], it is noted that for the carry chain lengths considered in Fig. 10, the local weak-indication self-timed system enables a 22% reduction in cycle time on average compared to the global weak-indication self-timed system.

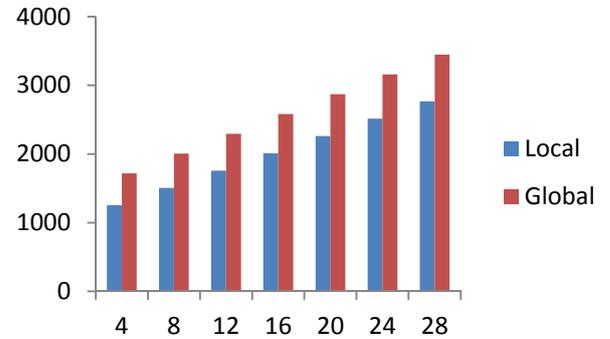

Fig. 10 Theoretical evaluation of cycle times of local and global weak-indication self-timed RCAs, assuming different carry propagation lengths
X-axis: Carry chain length; Y-axis: Cycle time in ns

## 4 Conclusions

Local and global weak-indication self-timed system architectures have been analyzed in our prior work [11] and it was found that global weak-indication helps to better optimize power, latency, and area compared to local weak-indication. This analysis has been extended in this work by considering the cycle time of the two system architectures since cycle time governs the throughput of a self-timed design. Whilst confirming the observations made in our previous work [11] that local weak-indication helps to better optimize power, latency and area compared to global weak-indication based on experimentation using an advanced 32/28nm CMOS process, it is additionally concluded in this work that local weak-indication is preferable with respect to cycle time optimization than global weak-indication. The final conclusions are: to achieve minimized area occupancy and power, global weak-indication is preferable, while to achieve better throughput, local weak-indication is preferable. Overall, from a combined power-cycle time-area consideration, local weak-indication may indeed be preferred. The conclusions derived in this work may be applicable for other adder architectures such as carry-lookahead and carry-select [39] – [42].

*References:*
[1] Available: http://www.itrs.net
[2] C.H. van Kees Berkel, M.B. Josephs, S.M. Nowick, "Scanning the technology: applications of asynchronous circuits," *Proc. the IEEE,* vo. 87, no. 2, pp. 223-233, February 1999.
[3] I. David, R. Ginosar, M. Yoeli, "Self-timed is self-checking," *Journal of Electronic Testing: Theory and Applications*, vol. 6, no. 2, pp. 219-228, April 1995.
[4] G.F. Bouesse, G. Sicard, A. Baixas, M. Renaudin, "Quasi delay insensitive asynchronous circuits for low EMI," *Proc. 4th International Workshop on Electromagnetic Compatibility of Integrated Circuits*, pp. 27-31, 2004.
[5] N.C. Paver, P. Day, C. Farnsworth, D.L. Jackson, W.A. Lien, J. Liu, "A low-power, low noise, configurable self-timed DSP," *Proc. 4th International Symposium on Advanced Research in Asynchronous Circuits and Systems*, pp. 32-42, 1998.
[6] K.J. Kulikowski, V. Venkataraman, Z. Wang, A. Taubin, M. Karpovsky, "Asynchronous balanced gates tolerant to interconnect variability," *Proc. IEEE International Symposium on Circuits and Systems*, pp. 3190-3193, 2008.
[7] I.J. Chang, S.P. Park, K. Roy, "Exploring asynchronous design techniques for process-tolerant and energy-efficient subthreshold operation," *IEEE Journal of Solid-State Circuits*, vol. 45, no. 2, pp. 401-410, February 2010.
[8] O.C. Akgun, J. Rodrigues, J. Sparsø, "Minimum-energy sub-threshold self-timed circuits: design methodology and a case study," *Proc. 16th IEEE International Symposium on Asynchronous Circuits and Systems*, pp. 41-51, 2010.
[9] Z.C. Yu, S.B. Furber, L.A. Plana, "An investigation into the security of self-timed circuits," *Proc. 9th International Symposium on Asynchronous Circuits and Systems*, pp. 206-215, 2003.
[10] D. Sokolov, J. Murphy, A. Bystrov, A. Yakovlev, "Design and analysis of dual-rail circuits for security applications," *IEEE Transactions on Computers*, vol. 54, no. 4, pp. 449-460, April 2005.
[11] P. Balasubramanian, N.E. Mastorakis, "Analyzing the impact of local and global indication on a self-timed system," *Proc. 5th European Computing Conference*, pp. 85-91, 2011.
[12] J. Sparsø, S. Furber, *Principles of Asynchronous Circuit Design: A Systems Perspective*, Kluwer Academic Publishers, Boston, MA, USA, 2001.
[13] P Balasubramanian, *Self-Timed Logic and the Design of Self-Timed Adders*, PhD thesis, School of Computer Science, The University of Manchester, 2010.
[14] T. Verhoeff, "Delay-insensitive codes – an overview," *Distributed Computing*, vol. 3, no. 1, pp. 1-8, March 1988.
[15] A.J. Martin, "The limitation to delay-insensitivity in asynchronous circuits," *Proc. 6th MIT Conference on Advanced Research in VLSI*, pp. 263-278, 1990.
[16] C.L. Seitz, "System Timing," in *Introduction to VLSI Systems*, C. Mead and L. Conway (Editors), pp. 218-262, Addison-Wesley, Reading, Massachusetts, USA, 1980.
[17] P. Balasubramanian, D.A. Edwards, "Efficient realization of strongly indicating function blocks," *Proc. IEEE Computer Society Annual Symposium on VLSI*, pp. 429-432, 2008.
[18] P. Balasubramanian, D.A. Edwards, "A new design technique for weakly indicating function blocks," *Proc. 11th IEEE Workshop on Design and Diagnostics of Electronic Circuits and Systems*, pp. 116-121, 2008.
[19] C.F. Brej, J.D. Garside, "Early output logic using anti-tokens," *Proc. 12th International Workshop on Logic and Synthesis*, pp. 302-309, 2003.
[20] P. Balasubramanian, "A robust asynchronous early output full adder," *WSEAS Transactions*

*on Circuits and Systems*, vol. 10, no. 7, pp. 221-230, July 2011.

[21] J. Sparsø, J. Staunstrup, "Delay-insensitive multi-ring structures," *Integration, the VLSI Journal*, vol. 15, no. 3, pp. 313-340, 1993.

[22] P. Balasubramanian, D.A. Edwards, "Self-timed realization of combinational logic," *Proc. 19th International Workshop on Logic and Synthesis*, pp. 55-62, 2010.

[23] P. Balasubramanian, R. Arisaka, H.R. Arabnia, "RB_DSOP: a rule based disjoint sum of products synthesis method," *Proc. 12th International Conference on Computer Design*, pp. 39-43, 2012.

[24] P. Balasubramanian, "A latency optimized biased implementation style weak-indication self-timed full adder," *Facta Universitatis, Series: Electronics and Energetics*, vol. 28, no. 4, pp. 657-671, December 2015.

[25] A.J. Martin, "Asynchronous datapaths and the design of an asynchronous adder," *Formal Methods in System Design*, vol. 1, no. 1, pp. 117-137, July 1992.

[26] C. Jeong, S.M. Nowick, "Block-level relaxation for timing-robust asynchronous circuits based on eager evaluation," *Proc. 14th IEEE International Symposium on Asynchronous Circuits and Systems*, pp. 95-104, 2008.

[27] P. Balasubramanian, R. Arisaka, "A set theory based factoring technique and its use for low power logic design," *International Journal of Computer, Electrical, Automation, Control and Information Engineering*, vol. 1, no. 3, pp. 721-731, 2007.

[28] P. Balasubramanian, D.A. Edwards, "A delay efficient robust self-timed full adder," *Proc. IEEE 3rd International Design and Test Workshop*, pp. 129-134, 2008.

[29] P. Balasubramanian, D.A. Edwards, W.B. Toms, "Redundant logic insertion and latency reduction in self-timed adders," *VLSI Design*, vol. 2012, Article ID 575389, pages 13, 2012.

[30] P. Balasubramanian, N.E. Mastorakis, "Timing analysis of quasi-delay-insensitive ripple carry adders – a mathematical study," *Proc. 3rd European Conference of Circuits Technology and Devices*, pp. 233-240, 2012.

[31] P. Balasubramanian, N.E. Mastorakis, "A set theory based method to derive network reliability expressions of complex system topologies," *Proc. Applied Computing Conference*, pp. 108-114, 2010.

[32] Synopsys Digital Standard Cell Library *SAED_EDK32/28_CORE Databook*, 2012.

[33] P. Balasubramanian, N.E. Mastorakis, "QDI decomposed DIMS method featuring homogeneous/heterogeneous data encoding," *Proc. International Conference on Computers, Digital Communications and Computing*, pp. 93-101, 2011.

[34] P. Balasubramanian, "Comments on "Dual-rail asynchronous logic multi-level implementation," *Integration, the VLSI Journal*, vol. 52, no. 1, pp. 34-40, January 2016.

[35] P. Balasubramanian, K. Prasad, N.E. Mastorakis, "Robust asynchronous implementation of Boolean functions on the basis of duality," *Proc. 14th WSEAS International Conference on Circuits*, pp. 37-43, 2010.

[36] P. Balasubramanian, D.A. Edwards, C. Brej, "Self-timed full adder designs based on hybrid input encoding," *Proc. 12th IEEE Symposium on Design and Diagnostics of Electronics Circuits and Systems*, pp. 56-61, 2009.

[37] P. Balasubramanian, D.A. Edwards, "Dual-sum single-carry self-timed adder designs," *Proc. IEEE Computer Society Annual Symposium on VLSI*, pp. 121-126, 2009.

[38] P. Balasubramanian, D.A. Edwards, "Heterogeneously encoded dual-bit self-timed adder," *Proc. 5th IEEE Conference on PhD Research in Microelectronics and Electronics*, pp. 120-123, 2009.

[39] P. Balasubramanian, D.A. Edwards, H.R. Arabnia, "Robust asynchronous carry lookahead adders," *Proc. 11th International Conference on Computer Design*, pp. 119-124, 2011.

[40] P. Balasubramanian, D.A. Edwards, W.B. Toms, "Self-timed section-carry based carry lookahead adders and the concept of alias logic," *Journal of Circuits, Systems and Computers*, vol. 22, no. 4, pp. 1350028:1-24, April 2013.

[41] P. Balasubramanian, D. Dhivyaa, J.P. Jayakirthika, P. Kaviyarasi, K. Prasad, "Low power self-timed carry lookahead adders," *Proc. 56th IEEE International Midwest Symposium on Circuits and Systems*, pp. 457-460, 2013.

[42] P. Balasubramanian, C. Jacob Prathap Raj, S. Anandhi, U. Bhavanidevi, N.E. Mastorakis, "Mathematical modeling of timing attributes of self-timed carry select adders," *Proc. 4th European Conference of Circuits Technology and Devices*, pp. 228-243, 2013.